\documentstyle[ip-author]{book}
\begin{document}
\title{Nonperturbative QCD Effects \\
         in Inclusive B Decays}
\author{Soo-Jong Rey}
{Physics Department \& Center for Theoretical Physics}
{Seoul National University, Seoul 151-742 KOREA \\
{\tt sjrey@phyb.snu.ac.kr}}
\abstract{
In this talk I report calculable nonperturbative QCD effects to the inclusive
semileptonic and radiative B-meson decay rates. Small instantons contribute to
the Wilson coefficient function at scales below $m_b$. It is found that the
instantons give rise to a potentially large correction to the decay rate near
the boundaries of the phase space for the semileptonic decay case, but totally
negligible for the radiative case. I explain how the difference may be
understood as a simple consequence of different kinematics in the two cases.
}

\def\be{\begin{equation}}
\def\ee{\end{equation}}
\def\bee{\begin{eqnarray}}
\def\eee{\end{eqnarray}}

\section{Introduction}
Weak decay of heavy flavors is an important probe to the fundamental
parameters in the Standard Model and to new physics beyond the
electroweak scale.
Of particular interests are inclusive semileptonic and radiative
decays of hadrons containing a single $b$ quark.
{}From the energy spectrum of charged lepton in $B \rightarrow X_q \ell
\overline \nu_\ell$
decay one may extract the heavy quark masses $m_b, m_c$ and the
Cabbibo-Kobayashi-Maskawa matrix elements $V_{qb}$ where $q = c, u$.
{}From the $B \rightarrow X_s \gamma$ decay one may probe the
indirect contribution of the heavy particles to the decay rate through the
penguin and Higgs exchange diagrams.
Traditional theoretical description of these inclusive decays was
based on models such as ACCMM model \bibref[accmm] or ISGW model
\bibref[isgw].

Recently Chay, Georgi and Grinstein \bibref[chayetal] have made an
important theoretical progress to the model-independent approach of the
inclusive $B$ decays (See also Ref. \bibref[bigietal].).
Using the heavy quark effective field theoy(HQEFT)
and the operator product expansion(OPE) Chay et.al. have shown that the
weak decay of $B$ mesons can be described by a systematic expansion
in powers of $ \Lambda / m_b$, of which the leading contribution
reproduces the parton-model result.

Using the OPE, Chay et.al. have shown that next-order
corrections are of order $(\Lambda /m_b)^2$,
where $\Lambda$ is the QCD scale, hence expected to be small.
Mannel~\bibref[mannel], Manohar and Wise~\bibref[manoharwise]
have analyzed the expansion systematically and concluded that these
corrections are singular at the boundaries
of the Dalitz plot, which necessitates some smearing prescription
before comparing with experimental data.

Experimentally it is necessary to make kinematical cuts to suppress
the background. In order to extract $V_{ub}$ from the shape of the
electron energy spectrum, one has to concentrate on the endpoint region
beyond the charm threshold. The available endpoint energy range is
${\cal O}(300 {\rm MeV})$, thus the final hadron is close to the
resonance region.
To extract a signal of new physics from the inclusive radiatve $B$ meson
decay, one has to concentrate on energetic photons near the kinematical
endpoint beyond the threshold of $K^* \rightarrow K \gamma$ secondary
decay. Again this is the resonance region of the final light hadrons.
In both cases the endpoint region is precisely the region at which
theoretical prediction is least understood due to various low-energy
QCD effects.

In this talk I report theoretical results on the calculable
nonperturbative QCD effects to the inclusive semileptonic and radiative
$B$ meson decays. In particular we calculate the effect of small QCD instantons
to these decay rates \bibref[chayrey,chayrey2].
In section 2 theoretical framework of calculating the decay rate is
summarized using the heavy quark effective field theory and
the operator product expansion.
In section 3 the instanton corrections are calculated.
Size of the correction is different for each decay processes.
In section 4 the origin of this difference is explained in terms
of the kinematics and the analytic structure of the forward Compton
scattering amplitude.
In section 5 measurement of invariant hadron mass distribution
is discussed.

\section{Operator Product Expansion for the Decay Rate}
The effective Hamiltonian of weak $B$ meson decay is obtained from
integrating out the massive $W$-boson, Higgs and heavy fermions
at the weak scale and scale down to $\mu = m_b$ using the renormalization
group equation
\be
H_{\rm eff} = \sum_{i} c_i(m_b) {\cal O}_i(m_b).
\ee
The relevant local operators for $B \rightarrow X_u l \nu_l$
and $B \rightarrow X_s \gamma$ are \bibref[inamilim,gsw]
\bee
{\cal O} &=& \bar u_L \gamma^\mu b_L \bar \nu_L \gamma_\mu \ell_L
\nonumber \\
{\cal O} &=& {e \over 16 \pi^2} F_{\mu \nu}
\bar s \sigma^{\mu \nu} (m_b P_+ + m_s P_-) b
\eee
respectively.
The decay rate of $B \rightarrow X + f$ ($f$ denotes the non-nonhadronic
final particles) is
\be
d \Gamma = {1 \over 2 m_b} \sum_{X} (2 \pi)^4
\delta^{(4)}(p_B - p_X - q)
|\langle X, f |c (\mu) \, {\cal O}(\mu) | B \rangle |^2
d ({\rm PS}),
\label{partonrate}
\ee
where $q = \sum_f p_f$ and
\bee
d ({\rm PS}) &=& \prod_f {d^3 \vec k_f \over (2 \pi)^3 2 E_f}
\nonumber \\
& = & {m_b^4 \over 2^7 \pi^4} d v \cdot \hat q \, d y \, d \hat q^2
\\
& = & {m_b^2 \over 4 \pi^2} v \cdot \hat q \, d (v \cdot \hat q)
\label{phasespace}
\eee
for semi-leptonic and radiative decays respectively. We have introduced
dimensionless kinematic variables $ \hat q = q / m_b$, $ y = 2 E_e / m_b$
and $\hat q^2 = q^2 / m_b^2$.

In the matrix element-squared in Eq.(3) the non-hadronic part $L$ is
trivially factorized. The remaining hadronic matrix element-squared
$W$ may be expressed in terms of the forward Compton scattering amplitude
using the optical theorem. Thus
\be
W (B \rightarrow X) = 2 {\rm Im} T (B \rightarrow X \rightarrow B)
\ee
where the forward Compton scattering amplitude is
\be
T = - i \int d^4 x e^{ - i q \cdot x}
\langle B| T\{ J^\dagger (x) J (0)\} | B \rangle
\ee

To evaluate the decay rate using the optical theorem it is necessasry to
examine the analytic structure of $T(q, v)$ \bibref[chayetal].
In the complex $v\cdot\hat q$ plane, the analytic structure for both
semi-leptonic and radiative decays are similar. For $B \rightarrow X \ell
\overline \nu_\ell$ the``physical cut'' relevant to the decay is located
on the real axis $v\cdot \hat q \le (1 + \hat q^2 -\rho)/2$  where
$\rho \equiv m_u^2 / m_b^2$ is the mass ratio-squared of the $u$ quark to
the $b$ quark.
For $B \rightarrow X_s \gamma$ the ``physical cut'' is located on the
real axis $ v \cdot \hat q \leq 1-r$ where $r = m_s^2/m_b^2$.
There are also another cut $  v \cdot \hat q \geq (3 - \hat q^2) / 2$ or
$(2+\sqrt{r})^2 -1 $ corresponding to other physical processes.
The analytic structures are shown in Fig.~1 for both cases.

The double differential decay rate after integrating over $ v \cdot \hat q$
is given by
\bee
{d^2 \Gamma \over d \hat q^2 \, d y}
&=& {G_F^2 m_b^5 \over 2^9 \pi^4} |V_{ub}|^2
\int_{{y \over 2} + {\hat q^2 \over 2 y}}^{{1 + \hat q^2 \over 2}}
\!\! d \hat v \cdot q \, { W \cdot  L \over m_b^2}
\nonumber \\
&=& - {G_F^2 m_b^5 \over 2^8 \pi^4} |V_{ub}|^2
 \int_C \! d v \cdot \hat q \,\, {T \cdot L \over m_b^2}
 \label{doublerate}
\eee
In the first equation the differential decay rate is evaluated from
the discontinuity of $T \cdot L$ along the contour $C^{\prime}$ in Fig.1.
In the second equation this is related to the contour integral
along the contour $C$ in the complex $v \cdot \hat q$ plane using
the Cauchy theorem. The contour $C$ is shown in Fig.1.

\begin{figure}
\vspace{2.2in}
\caption{\label{fig1a} analytic structure of $T$ relevant to the semileptonic
$B$ decay.}
\end{figure}

\begin{figure}
\vspace{2.2in}
\caption{\label{fig1b}
analytic structure of $T$ relevant to the radiative $B$ decay}
\end{figure}

Near the boundaries of the Dalitz plot $\hat q^2 \rightarrow y$ or
$y \rightarrow 1$ for semi-leptonic decay and $y_c = 1 - r$ for radiative
decay,
$z$ shrinks to zero and we are in the resonance region. There we expect
significant nonperturbative QCD effects.
This limits the extent of our theoretical prediction of the decay rate to
which the experimental data can be compared. Therefore it is
necessary to introduce a smearing in order to use perturbation theory
reliably~\bibref[pqw]. The size of the smearing is determined by requiring
that the nonperturbative effects be smaller than the parton-model result.
 Chay et.al. have observed that one can calculate the decay rate
 even in this limit perturbatively if it is possible to choose the integration
 contour away from the resonance region. The integral along $C$ is reliably
evaluated using perturbative QCD as long as $(m_b v - q)^2 \gg \Lambda^2$.

We have chosen the contour $C$ in Eq.~(\ref{doublerate}) as
a circle of radius $z  = (y - \hat q^2) ( 1 - y) / y$ centered at
$v \cdot \hat q = (1 + \hat q^2 )/2$ for $B \rightarrow X_u \ell
\overline \nu_\ell$ and a circle of radius $z = 2(1 -r - y_c)$
centered at $ v \cdot \hat q = 2(1 - r)$ for $B \rightarrow X_s \gamma$.
The soft photon cutoff $2 y_c$ is necessary experimentally to reduce
the background. The contribution of the contour integral near the point
$P$ is negligibly small as long as  $z \gg \Lambda / m_b$.

For the moment we consider $q^2 \ll m_b^2 $ but $(m_b v - q)^2 \gg
\Lambda^2$ in order to calculate the decay distribution reliably using
perturbative QCD. In the end we are interested in how far the result may be
extended to the endpoint region.
For processes involving $b$-quark decay it is appropriate to use the HQEFT.
In the HQEFT, the full QCD $b$ field is expressed in terms of
$b_v$ for $b$ quark velocity $v$. The $b_v$ field is defined by
\begin{equation}
b_v = \frac{1+ v \hskip -.215cm / }{2} e^{im_b v \cdot x} b + \cdots,
\end{equation}
where the ellipses denote terms suppressed by powers of $1/m_b$.
The equation of motion for $b_v$ is $v \hskip-0.215cm / b_v = b_v$.
We can apply the techniques of the OPE to expand $T$ in terms of matrix
elements of local operators involving $b_v$ fields
\begin{eqnarray}
\displaystyle
T
& = & -i \int d^4x e^{i q \cdot x}
\langle B | {\rm T} \{J(x)^{\dagger} J(0)\} | B \rangle
\nonumber  \\
& = & \sum_{n,v} C_n^{\mu\nu} (v, q) \,
\langle B | {\cal O}_v^{(n)} | B \rangle.
\label{ope}
\end{eqnarray}
Here ${\cal O}_v^{(n)}$ are local operators involving $\bar b_v b_v$ bilinears.
The coefficient functions $C^{\mu \nu}_n$ depend explicitly on
$q$ and $v$ because of the $v$ dependence of ${\cal O}_v^{(n)}$.
All large momenta are contained in the coefficient functions
while the matrix elements describe low-energy physics.

In this scheme $T$ is expanded as a double series
in powers of $\alpha_s$ and $1 / m_b$. To leading order in $\alpha_s$ and
$1 / m_b$ the forward Compton scattering amplitude for semi-leptonic and
radiative decays are
\begin{eqnarray}
T^s &=& -i \int \! d^4x \, e^{i {\cal Q} \cdot x} \,
 \langle B| T\{{\overline b}_v(x)
\gamma^{\mu}(1+\gamma_5) S_0 (x) \gamma^{\nu} (1+\gamma_5)b_v(0)\}
  |B\rangle
  \nonumber \\
T^r  &=& -i \int d^4 x e^{i {\cal Q} \cdot x} \,
  \langle B| {\overline b}_v (x) [k \hskip-0.22cm / , \gamma_\mu]
  (1 + \gamma_5) S_0(x)
  (1 - \gamma_5) [ k \hskip-0.22cm / , \gamma^\mu] b_v (0) | B \rangle
  \label{tstart}
  \end{eqnarray}
  \noindent
  respectively where ${\cal Q} = m_b v - k $ and $S_0(x)$ are the momentum
  and free propagator of the final quark.

When $ \Lambda^2 \ll {\cal Q}^2 \ll m_b^2$,
it is sufficient to keep the terms at leading order in
$1 / m_b$ only and expand Eq.~(\ref{tstart}) in powers of
$k / |{\cal Q}|$. This generates coefficient functions and
local operators in which $k$ is replaced by the derivatives
acting on the $b_v$ fields. The leading term independent of $k$ gives the
parton-model result. Contracting the leading term with $L$ we find
\begin{eqnarray}
T_0^s L & = &
128 m_b {k_e \cdot {\cal Q} k_{\bar \nu} \cdot v \over
{\cal Q}^2 };
 \nonumber \\
T_0^r &=& 16 m_b^3 (1 + r) {k \cdot {\cal Q} k \cdot v \over {\cal Q}^2}
\label{partonresult}
\end{eqnarray}
where $T_0$ is $T$ at leading order in $k/|{\cal Q}|$.

The next-order correction to  Eq.~(\ref{partonresult})
is suppressed by at least $(\Lambda / {\cal Q})^2$~\bibref[chayetal]
Corrections to $T$ from higher dimensional operators
involving $\overline b_v b_v$ bilinears have been studied
systematically~\bibref[bigietal,mannel,manoharwise,falketal,neubert].
Neubert~\bibref[neubert] has resummed these corrections to get a ``shape
function''. He has observed that the shape function is universal independent
of the final quark flavor and proposed to extract $V_{ub}$ from simultaneous
measurements of $b \rightarrow u$ semileptonic decay and radiative or $b
\rightarrow c$ semileptonic inclusive decays.

The coefficient functions beyond the leading order in $\alpha_s$ may be
calculated perturbatively. In addition there could be nonperturbative
correction to the coefficient functions~\bibref[shifman].
In the next section we calculate how instantons affect the coefficient
function of the leading term, Eq.~(\ref{partonresult}), in the OPE.
The instantons contribute not only to the matrix elements but
also to the coefficient functions.
We separate the instanton contributions to each of them as follows.
If there is an infrared divergence due to large instantons we
attribute it to the contribution to the matrix elements.
For infrared-finite parts we interpret them as the correction to
the coefficient functions.

\section{Calculable Instanton Correction}
We now compute the contribution of instantons to the coefficient
functions.
In estimating the contribution we start from the Euclidean
region where ${\cal Q}^2$ is large enough to use the OPE reliably.
We expect that the main contribution comes from small instantons
of size $\rho
\mathrel{\rlap{\lower3pt\hbox{\hskip0pt$\sim$}}\raise1pt\hbox{$<$}}
{1 /|{\cal Q}|}$, hence we use the dilute gas approximation
in what follows.
More specifically we calculate the instanton correction to the
decay rate at leading order in both $\alpha_s$ and $k$ in the OPE.
This is the instanton correction to the parton model result
Eq.~(\ref{partonresult}).

In the background of an instanton ($+$ anti-instanton) of size $\rho$
and instanton orientation $U$ located at the origin, the Euclidean
fermion propagator may be expanded in small fermion masses
as~\bibref[andreigross]
\begin{eqnarray}
\displaystyle
S_\pm (x, &y;& \rho_\pm; U_\pm)  =
- {1 \over m} \psi_0 (x) \psi^\dagger_0 (y)
+ S^{(1)}_\pm (x, y; \rho_\pm; U_\pm) \nonumber \\
&+& m \int d^4 w S^{(1)}_\pm (x, w; \rho_\pm; U_\pm)
S^{(1)}_\pm (w, y; \rho_\pm; U_\pm)
\nonumber \\
&+& {\cal O}(m^2),
\label{propexp}
\end{eqnarray}
where $\pm$ denotes instanton, anti-instanton.
$\psi_0$ is the fermion zero mode eigenfunction  and
$S^{(1)}_\pm = \sum_{E>0} {1 \over E} \Psi_{E \pm} (x)
\Psi^\dagger_{E \pm} (y)$ is the Green's function of fermion
nonzero modes.
In evaluating the forward Compton scattering amplitude
$T$,
we should use the propagator in Eq.~(\ref{propexp})
instead of $S_0$. $T^{\mu \nu}$ can be written as
\begin{equation}
T  = T_0  + T_{\rm inst.},
\label{tdecomp}
\end{equation}
where the first term is the parton-model amplitude.
The second term is the amplitude due to instantons of all
orientation $U$, position $z$ and size $\rho$.

After averaging over instanton orientations $T^{\mu \nu}_{\rm inst}$
is given by
\bee
T_{\rm inst.}  & = & \int \! d^4 \Delta \,
e^{i {\cal Q} \cdot \Delta } \sum_{a = \pm} \!\! \int
\! d^4 z_a \, d \rho_a D(\rho_a)
\nonumber \\
&\times & \langle B | \bar b_v (x) \gamma^\mu (1+\gamma_5) \{ {\cal S}_a
 (X, Y; \rho_a) -S_0 (\Delta)\} \gamma^\nu (1 + \gamma_5) b_v (y)
 | B \rangle,
 \label{insttmunu}
 \eee
 \noindent where $D(\rho)$ is the instanton density,
 $\Delta = x - y, \,\,\, X = x - z$ and $Y = y - z$.
 In Eq.~(\ref{insttmunu}), ${\cal S}_\pm (X, Y; \rho_\pm)$ is the
 fermion propagator averaged over instanton (anti-instanton)
 orientations centered at $z$.
 Using the ${\overline {\rm MS}}$ scheme with $n_f$
 flavors of light fermions, $D(\rho)$ is given by~\bibref[bernard]
 \begin{equation}
 \displaystyle
 D(\rho)  =  K \, \Lambda^5 \, (\rho \Lambda)^{6 + {n_f \over 3}}
  \, \biggl( \ln{1 \over \rho^2 \Lambda^2}
  \biggr)^{45-5 n_f \over 33 - 2 n_f},
  \label{density}
  \end{equation}
   where
  \be
 K   =  \biggl( \prod_i {\hat m_i \over \Lambda} \biggr) \,
 2^{12 n_f \over 33 - 2n_f} \, \biggl({33 - 2n_f \over 12} \biggr)^6
  {2 \over \pi^2} \, \exp\bigl[{1 \over 2} - \alpha (1) + 2 (n_f -1)
 \alpha ({1 \over 2}) \bigr]
 \label{ddensity}
 \ee
  in which the $\beta$ function at two loops and the running mass
 at one loop are used and $\hat m_i$ are the
renormalization-invariant quark masses.
In Eq.~(\ref{ddensity}) $\alpha(1) = 0.443307$ and
$\alpha(1/2) = 0.145873$. From now on we replace the logarithmic
term in $D(\rho)$ by its value for $\rho = 1/|{\cal Q}|$.
Corrections to this replacement are
 negligible as they are logarithmically suppressed.

Inserting Eq.~(\ref{propexp}) to Eq.~(\ref{insttmunu}),
the leading contribution comes from the mass-independent part
${\cal S}_\pm^{(1)}$ due to the chiral structure of the
left-handed weak currents.
Keeping the chirality-conserving part of ${\cal S}^{(1)}_\pm$ in the singular
gauge~\bibref[andreigross] $T_{\rm inst}$ is evaluated straightforwardly. The
$z$ integral is convergent as $|z| \rightarrow \infty$.
The integration over $\rho$ is convergent for small instantons.
On the other hand large instanton part is divergent.
However since the integrand is analytic for large $\rho$, there are only a
finite number of divergent terms when $T^{\mu \nu}_{\rm inst}$ is expanded in
$1 / \rho$.  We interpret these infrared divergent terms as the instanton
 contribution to the matrix elements of operators in the OPE~\bibref[balietal].
 The remaining terms are infrared convergent and are interpreted that they
 contribute to the coefficient functions. In order to calculate finite
terms any convenient regularization prescription will do.
To this end we analytically continue the exponent of $\rho$ in the
instanton density so that $D(\rho) \propto
\rho^{M-4}$~\bibref[porrati] and the spacetime dimensions to
$d = 4 + 2 \epsilon$.
In the end we let $M \rightarrow 11$ for $n_f = 3$ and
$\epsilon \rightarrow 0$ to get the final answer.

 It is now straightforward to evaluate the instanton contribution
 to the differential decay rate.
 So far we have worked in Euclidean spacetime.
 We now make a naive analytic continuation to Minkowski spacetime
 with timelike ${\cal Q}^2 = (m_b v - q)^2$ to evaluate the
 differential decay rate.
  For the semileptonic decay, the result of \bibref[chayrey] for the
  double differential decay rate is
  \begin{equation}
  \displaystyle
  {1 \over \Gamma_0} {d^2 \Gamma_{\rm inst} \over d \hat q^2 d y} =
  A \, y^5 \,
  { 5 \hat q^2 - (1-y) (y-\hat q^2) \over (1 - y)^{6}
  (y - \hat q^2)^{5}}.
  \label{ddrate2}
  \end{equation}
  where
  \be
  A = {2^{11} \pi^2 \over 175} \Gamma^2(6) K \bigg(
   {\Lambda \over m_b}\biggr)^{12}.
  \ee
  The instanton effect is suppressed at $y \sim 0$.
  However it become singular near the boundaries of the phase space
  at $y \sim 1$ and $y \sim \hat q^2$.
  At these boundaries the final quark approaches the resonance region.
  As there are large instanton contributions near the resonance point,
  it is necessary to introduce a smearing to define sensible
  decay rate in perturbative QCD.

  As a simple prescription we first consider the smeared single
  differential decay rate
  \begin{equation}
  \langle {d \Gamma_{\rm inst} \over d y} \rangle_\delta
  = \int_0^y d \hat q^2 \, \theta( y - \hat q^2 - \delta)
  {d \Gamma_{\rm inst} \over d y}
  \label{singsmear}
  \end{equation}
  in which we restrict the phase space so that the singular region
  $y = \hat q^2$ is avoided by $\delta$.
  The size of smearing $\delta$ is determined by the requirement
  that the smeared instanton contribution to
  the single differential decay rate be smaller than that in
  the parton model
  $d \Gamma_0 / d y = 2 y^2 (3 - 2 y) \Gamma_0$.
  Eq.(\ref{singsmear}) is given by
  \begin{eqnarray}
  \langle {d \Gamma_{\rm inst} \over d y} \rangle_\delta
  & = & \bigl({d \Gamma_0 \over d y} \big)
  { A \over 24} {1 \over (1-y)^6 (3 - 2y)}
  \nonumber \\
  & \times & \bigl[9 - 4 y - 24({y \over \delta})^3 + 15
  ({y \over \delta})^4 + 4 ({y^4 \over \delta^3}) \bigr].
  \label{singsmear2}
  \end{eqnarray}

  The ratio $R(y, \delta) = \langle d \Gamma_{\rm inst}
  / dy \rangle_\delta / (d \Gamma_0 / dy)$
  is plotted in Fig.2 for three different values of
  $\delta$ = 0.15, 0.17, 0.19 with $\Lambda = 400$ MeV and
  $m_b = 5$ GeV.
  \begin{figure}
  \vspace{3.5in}
\caption{\label{fig2}
$R(y,\delta)$ for $\delta$ = 0.15, 0.17, 0.19 with
$\Lambda = 400 $MeV and $m_b = $ 5 GeV.}
\end{figure}

  The numerical value of $A$ is given by
  \begin{equation}
  A = \biggl({19.2 {\rm GeV} \over m_b} \biggr)^3 \,
  \biggl({\Lambda \over m_b} \biggr)^9,
  \label{numa}
  \end{equation}
  in which we set $n_f = 3$ and the renormalization-invariant
  quark masses as~\bibref[quarkmass]
    $\hat m_u  = 8.2 \pm 1.5 {\rm MeV}, \hat m_d  =  14.4 \pm 1.5 {\rm MeV},
    \hat m_s  =  288 \pm 48 {\rm MeV}$.
   We see that for these choices of $\delta$, $R (y, \delta)$
   grows rapidly for $y
   \mathrel{\rlap{\lower4pt\hbox{\hskip1pt$\sim$}}\raise1pt\hbox{$>$}}
   0.84$. From Fig.2 we see that a smearing of
   $ \delta \approx 0.16 \sim 0.20$ at the boundary
   $y = \hat q^2$ of the phase space is needed to extract a
   reliable electron spectrum from the single differential decay rate.

       We have also examined the behavior of $R(y, \delta)$ as we increase
       $\delta$.
       The position at which $R(y, \delta)$ grows like a brick wall
       does not shift much from $y \sim 0.84$. Because of this singular
       behavior we need another smearing near $y = 1$.
       For simplicity we cut off the region near $y = 1$ by the
       same smearing width $\delta$ as in Eq.~(\ref{singsmear}).

       Instanton contribution to the total decay rate after such a smearing
       prescription is given by
       \begin{eqnarray}
       \displaystyle
       \langle  \Gamma_{\rm inst} \rangle_\delta & =&
       \! \int_0^1 \! d y \! \int_0^y \! d \hat q^2
       \theta(y-\hat q^2 -\delta) \theta(1 - y - \delta)
       {d^2 \Gamma_{\rm inst} \over d \hat q^2 d y} \nonumber \\
       & = &  \Gamma_0 \, {A \over 24}
       \, {1 \over \delta^9}
       ( \, 6 - 53 \delta + 198 \delta^2 - 420 \delta^3
       + 612 \delta^4  \nonumber \\
       && - 354 \delta^5 + 16 \delta^6 - 4 \delta^7 - \delta^9
       + 180 \delta^5 \ln \delta).
       \label{totsmear}
       \end{eqnarray}
       The ratio $R(\delta) = \langle \Gamma \rangle_\delta / \Gamma_0$
       is plotted in Fig.3 for three different values of $\Lambda =
       350, 400, 450$ MeV and for $m_b = 5$ GeV.

       \begin{figure}
       \vspace{3.5in}
\caption{\label{fig3}
$R(\delta)$ for $\Lambda = $ 350, 400, 450 MeV with $m_b = $ 5 GeV.}
       \end{figure}

       We see that the ratio $R (\delta)$ also rises sharply like
       a brick wall as $\delta$ decreases.
       When $\delta$ is large, say $\delta
       \mathrel{\rlap{\lower4pt \hbox{\hskip1pt$\sim$}}\raise1pt\hbox{$>$}}
       0.15$, $R (\delta)$ is
       insensitive to the choice of $\Lambda / m_b$. However, for small
       $\delta$, say $ \delta \approx 0.12$, $R(\delta)$ is sensitive to
       the value of $\Lambda / m_b$. If we require $R (\delta )$ be less
       than 20\%,
       the size of the smearing is roughly at least $ 0.12 \sim 0.15$
       for our choices of $\Lambda$.

We have also calculated the instanton effect for the inclusive radiative
decay rate.
  Instanton effects on the $B \rightarrow X_s\gamma$ decay depend on the
momentum cutoff of the final-state photon. The cutoff to remove soft
photons should be introduced in experiments since it is difficult to
isolate a soft photon from the background coming
from the subsequent decay of, say, $K^* \rightarrow K + \gamma$.
Experimentally the photon spectrum from $B \rightarrow X_s\gamma$ is
concentrated in the region $2.2\ {\rm GeV} \le E_{\gamma}
\le 2.5
\ {\rm GeV}$. In CLEO the lower cut is taken at $ E_\gamma \sim 2.2
$GeV ~\bibref[cleo]. Thus we restrict the
region of contour integral to $y_c \le v \cdot \hat q \le 1$ where $y_c$ is
the experimental cutoff for the scaled photon energy.
Then the instanton decay rate is~\bibref[chayrey2]
\begin{equation}
\frac{\Gamma_{\rm inst} (y_c)}{\Gamma_0}
= \biggl(\frac{6.67 \mbox{GeV}}{m_b} \biggr)^3
\ \biggl({\Lambda \over m_b} \biggr)^{9} \
\biggl[\frac{10}{(1-y_c)^6}-\frac{36}{(1-y_c)^5}+ \frac{45}{(1-y_c)^4}-
\frac{20}{(1-y_c)^3}\biggr].
\end{equation}
The ratio $\Gamma_{\rm inst}(y_c)/\Gamma_0$ is shown in Fig.~4 as a
function of the cutoff $y_c$ for different values of
$\Lambda = 350, 400, 450 $ MeV with $m_b = 5$ GeV.
\begin{figure}
\vspace{3.5in}
\caption{\label{fig4}
$\Gamma_{\rm inst}(y_c) / \Gamma_0$ for $\Lambda = $ 350, 400, 450 MeV
with $m_b$ = 5 GeV.}
\end{figure}
For $y_c \approx 0.82$ that CLEO has chosen, the instanton correction
$13.1 \times 10^{-4}$ as $\Lambda$ varies from $350$ MeV
to $450$ MeV. Therefore the instanton correction is negligibly small
compared to other corrections.
Note that the instanton contribution is much smaller as the cutoff
$y_c$ decreases.
On the other hand, as Fig.4 shows, the contribution is appreciable
at $y_c \ge 0.92$. In this region we need a smearing to make the
perturbation theory valid.
However this cutoff is too large to be significant experimentally
since there is a very small fraction of the rate in this energy window.

The $\alpha_s$-correction to the decay rate also becomes large as $y_c$
approaches $1$. Ali and Grueb ~\bibref[aligrueb] have examined the
correction in detail and have concluded that
the $\alpha_s$-correction has to be exponentiated near the endpoint
of the photon spectrum.
Numerically they have found that the exponentiation is necessary for
$y_c \ge 0.85$.
Combined with their result our analysis indicates that there is
a large theoretical uncertainty for the cutoff $y_c \ge 0.88$.
Therefore in order to compare experiments with theory in a
model-independent way the cutoff $y_c$ has to be chosen below 0.88.

We also obtain the instanton contribution to the differential decay
rate as
\begin{equation}
\frac{d \Gamma_{\rm inst}}{d y} = \Gamma_0 \frac{N_{\rm inst}}{m_b^{12}}
{y^3 \over (1 - y)^7}.
\end{equation}
Numerically we find
\begin{equation}
\frac{1}{\Gamma_0} \frac{d \Gamma_{\rm inst}}{d y}
= \biggl(\frac{26.1\ \mbox{GeV}}{m_b} \biggr)^3
\bigl(\frac{\Lambda}{m_b} \bigr)^9 {y^3 \over (1 - y)^7}.
\label{diffratio}
\end{equation}
The ratio in Eq.~(\ref{diffratio}) is shown in Fig.5 for different values
of $\Lambda = 350, 400, 450 $ GeV with $m_b = 5$ GeV.
\begin{figure}
\vspace{3.5in}
\caption{\label{fig5}
$(1/\Gamma_0)d\Gamma_{\rm inst}/dy$ for $\Lambda=$ 350, 400, 450 MeV with
$m_b$ = 5 GeV.}
\end{figure}
The ratio also increases sharply near the endpoint of the photon spectrum.
Above $y \approx 0.87$ the instanton correction is appreciable
and necessitates a smearing in this region.

\section{Kinematics and Analyticity}
We have found that the instanton effect is quite different for
different inclusive decay modes.
For $B \rightarrow X_s \gamma$, the effect to the total decay rate
is negligible as long as $y_c$ is small enough.
This is in contrast to the case of $B \rightarrow X_u e \overline
\nu_e$ decay in which the differential decay rate receives a large
instanton correction. This apparently big difference can be understood from
kinematics.

As explained in Section 2 we have obtained the averaged total decay rate
by deforming the integration contour (from $C'$ to $C$ in Fig.1).
If the deformed contour is chosen sufficiently far away from the
resonance region we can calculate the decay rate reliably using the
perturbation theory. It depends on the details of the kinematics whether
such contour deformation is possible.

In $B\rightarrow X_s\gamma$ decay the radius of the deformed contour $C$
is fixed at $1 - y_c$. This radius represents the off-shell
invariant mass-squared of the final hadrons.
Note that this radius is independent of kinematic variables.
As long as $1 - y_c \gg \Lambda^2 / m_b^2$ the averaged decay rate
can be calculated reliably at the scale $m_b^2 (1 - y_c)$.
The instanton contribution is evaluated at this scale.
As calculated in Section 3, the contribution is ${\cal O}(10^{-4})$.

In $B \rightarrow X_u e \overline \nu_e$ decay the radius of the
deformed contour is given by $z = (y - \hat q^2)(1 - y) / y$ where
$ y = 2 E_e /m_b$ and $\hat q^2 = (k_e + k_\nu)^2 / m_b^2$ ~\bibref[chayrey].
The off-shell invariant mass-squared of the final hadrons is equal
to $m_b^2 z$. As long as $z \gg \Lambda^2 / m_b^2$ we can calculate
the averaged decay rate using perturbation theory. However unlike the
$B\rightarrow X_s\gamma$ case the radius $z$ depends on how the momentum
transferred to the leptonic system is distributed between the electron
and the anti-neutrino.
In particular $z$ vanishes at the boundaries of the phase space.
Therefore the instanton contribution evaluated at the scale $m_b^2 z$
grows rapidly near the boundaries $y = 1$ and $y= \hat q^2$.
This means that it is impossible to avoid the resonance region unless we
introduce a model-dependent cut near the boundaries of the phase space.

Note that the dependence of the off-shell invariant mass-squared of
the final hadrons on the kinematic variables arises only when there
are more than one ``nonhadronic'' particle in the final state.
In this respect it is interesting to compare our result with the
inclusive hadronic $\tau$ decay $\tau \rightarrow \nu_{\tau} +X$
{}~\bibref[braaten]. In this case the neutrino plays the same role as the
photon in the $B\rightarrow X_s\gamma$ decay as far as the kinematics
is concerned. The maximum invariant mass-squared $s = q^2$ of the final
hadrons is fixed at $m_{\tau}^2$ independent of the neutrino energy.
Since this is much larger than $\Lambda^2$ the total inclusive decay
rate can be calculated reliably using the deformed contour.
While the specific kinematic variables under consideration are different,
the analytic structure and the fact that the maximal invariant
mass-squared of the final hadrons is independent of other kinematic
variables are similar in both $B \rightarrow X_s \gamma$ and inclusive
hadronic $\tau$ decays.

Indeed the instanton contributions are small in both cases. Nason and
Porrati~\bibref[porrati] have obtained the instanton contribution to the
ratio of the hadronic to the leptonic width $R_{\tau}$ which is given
by
\begin{equation}
\frac{R_{\tau}^{\rm inst}}{R_{\tau}^0} =
\biggl(\frac{3.64\Lambda}{m_{\tau}}\biggr)^9 \frac{{\hat m}_u {\hat m}_d
{\hat m}_s}{m_{\tau}^3}.
\end{equation}
Our result for the instanton contribution to the decay rate of
$B \rightarrow X_s \gamma$ is written as
\begin{equation}
\frac{\Gamma_{\rm inst}}{\Gamma_0} =
\biggl(\frac{5.91\Lambda}{m_b}\biggr)^9 \frac{{\hat m}_u {\hat m}_d
{\hat m}_s}{m_b^3}
\end{equation}
for $y_c = 0$. The fact that both have the same mass dependence and
similar numerical factors confirms our expectation.

By the same argument we expect the instanton contribution to the
$B \rightarrow X_s e^+ e^-$ decay rate is similar to that of the
$B \rightarrow X_u e \overline \nu_e$ because kinematics
and the analytic structure of the forward Compton scattering amplitudes
are the same.

\section{Discussions}
In this talk I have reported on nonperturbative QCD effects to the
inclusive semileptonic and radiative $B$-meson decays. The final
hadrons are on the mass-shell at the boundaries of the phase space.
If the final hadron has an invariant mass below the QCD scale, then
the nonperturbative QCD effects affect not only the hadron matrix elements
but also the coefficient functions.
Our calculation is consistent with what we expect from the consideration of the
analyticity and the kinematics of each decay channels.
The instanton effect is quite sizable for semileptonic decays but
negligibly small for the radiative decay.

While the model-dependence seems inevitable to extract out $V_{ub}$ from
the shape of the electron spectrum in $B \rightarrow X_u e \bar \nu_e$,
one may alternatively measure the shape of the inclusive hadron energy
spectrum.
This is not possible at present but the next generation $B$-factories as
are designed definitely offer such a possibility.
In this case the previous analysis based on ACCMM model~\bibref[barger]
indicates that there is a rather wide $\sim 1.5 $ GeV range at which the
$V_{ub}$ may be safely extracted without encountering the resonance region.
Model-independent QCD analysis to this proposal is currently under
investigation
and hope to report at the next meeting.

I thank J.G. Chay for discussions and collaboration. This work was
supported in part by Grants from KOSEF-SRC program, Ministry of Education
BSRI 94-2108 and Korea Research Foundation Nondirected Research Grant '93.

\begin{bibliography}

\bibitem[accmm]
{G. Altarelli, N. Cabibbo, G. Corbo, L. Maiani and G. Martinelli,
Nucl. Phys. \bf B208 \rm, 365 (1982). }

\bibitem[isgw] {B. Grinstein, N. Isgur and M.B. Wise,
Phys. Rev. Lett. \bf 56 \rm, 258 (1986);
N. Isgur, D. Scora, B. Grinstein and M.B. Wise,
Phys. Rev. \bf D39 \rm, 799 (1989).}

\bibitem[chayetal]{ J. Chay, H. Georgi and B. Grinstein,
Phys. Lett. \bf 247B \rm, 399 (1990).}

\bibitem[bigietal]{
I.I. Bigi, M.A. Shifman, N.G. Uraltsev and A.I. Vainshtein,
Phys. Rev. Lett. \bf 71 \rm, 496 (1993);
B. Blok, L. Koyrakh, M. Shifman and A.I. Vainshtein,
Phys. Rev. \bf D49 \rm (1994) 3356.  }

\bibitem[mannel]{ T. Mannel, Nucl. Phys. \bf B413 \rm, 396 (1994).
}

\bibitem[manoharwise]{ A.V. Manohar and M.B. Wise,
Phys. Rev. \bf D49 \rm (1994) 1310. }

\bibitem[chayrey]{ J. Chay and S.-J. Rey,
SNUTP 94-08 preprint, 1994 (unpublished).
}

\bibitem[chayrey2]{ J. Chay and S.-J. Rey,
SNUTP 94-54 preprint, 1994 (unpublished).
}

\bibitem[inamilim]{
 T. Inami and C.S. Lim, Prog. Theor. Phys.
 \bf 65 \rm (1981) 297.
 }

 \bibitem[gsw]{ B. Grinstein, R. Springer and M.B. Wise, Phys. Lett. \bf 202B
 \rm (1988) 138; B. Grinstein and M.B. Wise,
 Phys. Lett. \bf B201 \rm (1988) 274.
 }

\bibitem[bigiope]{ I.I. Bigi, M.A. Shifman, N.G. Uraltsev and
A.I. Vainshtein, TPI-MINN-93/60-T preprint, 1993 (unpublished).
}

\bibitem[pqw]{ E.C. Poggio, H.R. Quinn and S. Weinberg,
Phys. Rev. \bf D13 \rm, 1958 (1976).
}

\bibitem[falketal]{
A. Falk, M. Luke and M.J. Savage, UCSD/PTH 93-23 preprint, 1993
(unpublished).
}

\bibitem[neubert]{ M. Neubert,
CERN-TH.7087/93 preprint, 1993 (unpublished).
}

\bibitem[shifman]{
V.A. Novikov, M.A. Shifman, A.I. Vainshtein and V.I. Zakharov,
Nucl. Phys. \bf B174 \rm, 378 (1980).
}

\bibitem[andreigross]{ N. Andrei and D. Gross,
Phys. Rev. \bf D18 \rm, 468 (1978).
}
\bibitem[bernard]{ G. 't Hooft, Phys. Rev. \bf D14 \rm, 3432 (1976);
G. 't Hooft, Phys. Rep. \bf 142\rm, 357 (1986);
C. Bernard, Phys. Rev. \bf D19 \rm, 3013 (1979).
}
\bibitem[balietal]{
L. Balieu, J. Ellis, M.K. Gaillard and W.J. Zakrzewski,
Phys. Lett. \bf 77B \rm, 290 (1978).
}

\bibitem[porrati]{
I.I. Balitskii, M. Beneke and V.M. Braun, Phys. Lett.
\bf 318B \rm, 371 (1993);
P. Nason and M. Porrati,
CERN-TH.6787/93 preprint, 1993 (unpublished).
}
\bibitem[quarkmass]{ C.A. Dominguez and E. de Rafael,
Ann. Phys. \bf 174 \rm, 372 (1987).
}

\bibitem[cleo]{
For a comprehensive review, see T.E. Browder, K. Honscheid and S. Playfer,
CLSN 93/1261, HEPSY 93-10 preprint, 1993 (unpublished).
}

\bibitem[aligrueb]{ A. Ali and C. Grueb, Phys. Lett. \bf B287 \rm (1992) 191.
}

\bibitem[braaten]{ E. Braaten, S. Narison and A. Pich,
Nucl. Phys. \bf B373 \rm (1992) 581.
}

\bibitem[barger]{V. Barger, C.S. Kim and R.N. Phillip, Phys. Lett. \bf B251 \rm
(1992) 629.
}

\end{bibliography}
\end{document}